\documentclass[3p,11pt]{elsarticle}

\usepackage{amssymb}
\usepackage{hyperref}
\usepackage{rotating}

\newcounter{bla}

\journal{Computer Physics Communications}

\begin{document}

\begin{frontmatter}



\title{Electron Transport in Gaseous Detectors with a {\tt Python}-based Monte Carlo Simulation Code}

%
\author[1]{B.~Al Atoum }
\author[2]{S. F. Biagi}
\author[3]{D.~Gonz\'alez-D\'iaz}
\author[1]{B.J.P.~Jones}
\author[1]{A.D.~McDonald}

\cortext[author] {Corresponding author.\\\textit{E-mail address:} bashar.atoum@mavs.uta.edu}

\address[1]{
Department of Physics, University of Texas at Arlington, Arlington, TX 76019, USA}

\address[2]{
University of Liverpool, Physics Department, Liverpool L69 7ZE, United Kingdom}

\address[3]{
Instituto Galego de F\'isica de Altas Enerx\'ias, Univ.\ de Santiago de Compostela, Campus sur, R\'ua Xos\'e Mar\'ia Su\'arez N\'u\~nez, s/n, Santiago de Compostela, E-15782, Spain}

\begin{abstract}
Understanding electron drift and diffusion in gases and gas mixtures is a topic of central importance for the development of modern particle detection instrumentation.  The industry-standard {\tt MagBoltz} code has become an invaluable tool during its 20 years of development, providing capability to solve for electron transport (`swarm') properties based on a growing encyclopedia of built-in collision cross sections.  We have made a refactorization of this code from {\tt FORTRAN} into {\tt Cython}, and studied a range of gas mixtures of interest in high energy and nuclear physics. The results from the new open source {\tt PyBoltz} package match the outputs from the original {\tt MagBoltz} code, with comparable simulation speed. An extension to the capabilities of the original code is demonstrated, in implementation of a new Modified Effective Range Theory interface.  We hope that the versatility afforded by the new {\tt Python} code-base will encourage continued use and development of the {\tt MagBoltz} tools by the particle physics community.
\end{abstract}

\end{frontmatter}






\section{Introduction}
\label{intro}
The development of software that can accurately describe the transport properties of electrons in gas has been invaluable in the development and design of modern gaseous detectors.  Experiments based on devices such as time projection chambers, drift chambers, and multiwire or micropattern proportional chambers rely critically on the realization of gas mixtures that optimize various figures of merit including  charge multiplication and scintillation, attachment, diffusion or mobility~\cite{sauli2014gaseous,Gonzalez-Diaz:2017gxo}.  These properties can, under suitable assumptions, be calculated based on measured or swarm-parameter-based collision cross sections via Monte Carlo codes.  Several software packages are presently available~\cite{Veenhof:2009zza} each with somewhat different applications and approaches. Among the more prominent are {\tt MagBoltz}~\cite{biagi1999monte}, its sister-code {\tt Degrad}, and {\tt Garfield++}~\cite{Garfield} (which also uses {\tt MagBoltz} cross sections), as well as others with more localized user bases. Many of the codes track the properties of an electron swarm that is evolving in time in step-wise manner, sampling from collision cross sections to evolve the ensemble in phase space. Given accurately described cross sections, theses packages can provide critical information on electron drift in gas mixtures. 

{\tt MagBoltz}, used either directly or with its cross sections interfaced by {\tt Degrad} or {\tt Garfield++}, is one of the most widely used electron swarm simulation codes (a handful of applications include, for example, Refs.~\cite{Ruiz-Choliz:2015daa,Azevedo:2014hka,riegler2003detector,Burns:2017dny,Sahin:2018ybn,Simon:2018vep}). It is written in {\tt FORTRAN}, with a built-in library of collision cross sections that is evolving continuously as the necessity for more accurate transport parameters or the availability of new gases dictates.  This package is world-leading in terms of comprehensiveness of the cross section library and performance.  Implementation within {\tt FORTRAN}, however, implies some practical limitations that can represent a barrier against inclusion of new functionalities, complicate interfaces to other codes, and discourage some developers from working with the code-base. Students and Postdocs in High Energy and Nuclear Physics today are typically fluent in C++ and {\tt Python}, for example, but infrequently expert at {\tt FORTRAN}.

Motivated by an interest in studying the properties of diffusion-reducing gas mixtures for neutrinoless double beta decay~\cite{Henriques:2017rlj,Henriques:2018tam,Azevedo:2015eok}, we have undertaken a re-factorization of the original {\tt MagBoltz} code into a more modern language.  Our past use cases of the original {\tt MagBoltz} code have included making systematic explorations of xenon-based gas mixtures for reduced transverse diffusion~\cite{XePa}.  Helium appears to be an especially promising additive, and was studied using {\tt MagBoltz} simulations in Ref.~\cite{Felkai:2017oeq}. The mixture has now been tested experimentally both in terms of its electron-cooling properties~\cite{McDonald:2019fhy}, and electroluminescence light yield~\cite{Fernandes:2019zuz}.  A continuing experimental program with xenon/helium mixtures is under way to establish the effect on the topological signature of $0\nu\beta\beta$ within the NEXT-DEMO++ program~\cite{NeusLIDINE}. 

Ongoing efforts to understand the detailed microscopic behaviour of electrons in various gas mixtures, including but not limited to diffusion suppression in xenon+helium, has required studying and modifying the {\tt MagBoltz} calculation in some detail.  This prompted us to re-factorize the original {\tt FORTRAN} code into a more flexible format.  Our refactorization involved a near-complete rewrite, redesigning to incorporate a modular and object-based structure, and re-optimizing the program flow.  Algorithmically, the calculations are equivalent to the modern version of {\tt MagBoltz}, and we take this opportunity to unambiguously assign all scientific credit for algorithmic development, tuning and evolution to original author, Steve Biagi~\cite{biagi1999monte}. 

The framework chosen to support this project is the {\tt Cython} \cite{behnel2011cython} extension of {\tt Python}. {\tt Cython} maintains the flexibility and code syntax of {\tt Python} while inheriting some functionality from C++ to allow compilation, for improved speed of numerical calculations.  The choice of {\tt Cython} reflects the combined goals of implementing a {\tt Python}-style interface for ease of use while maintaining the computational performance of the  lower-level {\tt FORTRAN} language (Sec.~\ref{ PyBoltz }).  The new {\tt PyBoltz} code and documentation is publicly available at~\cite{GitHub}, and is provided as open source, with further development and extension encouraged.

\section{Electron Transport Implementation}
\label{Theory}
The original {\tt MagBoltz} code obtained its name on the basis of being a solver of the {\tt Boltz}mann equation in a {\tt Mag}netic field~\cite{biagi1988accurate,biagi1989multiterm}. However, since 1999, calculations within {\tt MagBoltz}  have been based instead upon Monte Carlo integration, following approximately the methods of Frasier and Mathieson~\cite{fraser1986monte}. The {\tt PyBoltz} code utilizes the same Monte-Carlo integration technique as {\tt Magboltz}, which was outlined by Biagi in~\cite{biagi1999monte}.  Here we describe this  method.

For the purposes of optimal computation speed, independent integrators are implemented for transport with and without thermal motion, and with no magnetic field, magnetic field parallel to the electric field, and with magnetic field at a generic angle to electric field.   

The simulation proceeds electron-by-electron, and collision-by-collision.  A specified number of real collisions are simulated, divided into a small number of ``samples''. The samples each provide an independent measure of all drift parameters, and their standard deviation is used to assess statistical uncertainty on the simulation.  Presently, all transport parameters are extracted from a single electron, propagated over a sufficiently large number of collisions that it is assumed to ergodically explore the available configuration space.

As pointed out by Skullerud \cite{skullerud1968stochastic}, sampling of time-to-next-collision given a general velocity-dependent cross section in principle involves a costly numerical integration between each scattering event.  This computational problem can be overcome via Skullerud's Null Collision method.  Here, collisions are forced at a frequency much higher than the true collision frequency. However, the majority of these collisions are ``Null'' collisions, in that they transfer no energy or momentum. The benefit of this method is that the time between collisions is forced to be sufficiently short that the collision cross section can be assumed to be locally velocity independent. In such cases, the kinematic equations for electron transport can be solved analytically to yield the probability distribution for time-to-next-null-collision.  This distribution is independent of applied magnetic field since it does not affect the energy of the electron in flight.  After an appropriate, randomly sampled number of null collisions,  a real collision is forced at a frequency determined by the various scattering cross sections of the gases.  This method offers a substantial performance improvement over calculation of the time-to-next-real-collision directly.

\begin{figure}
\begin{centering}
\includegraphics[width=0.99\columnwidth]{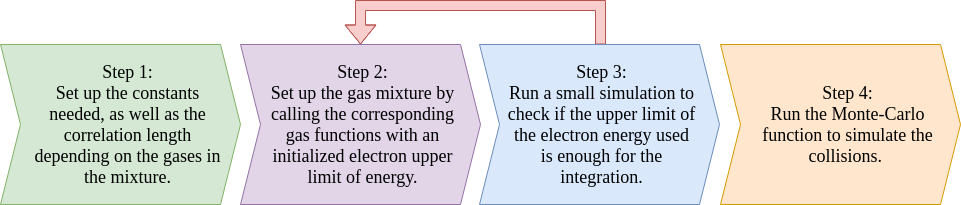}
\par\end{centering}
\caption{Simplified flow charge showing the PyBoltz / MagBoltz method of estimating the upper energy limit for binned evolution.
\label{fig:PyBoltz_Flow}}
\end{figure}

Before simulations begin, the gas properties are processed to produce an effective summed cross section for each gas. Data tables of the elastic, elastic momentum transfer, attachment, rotation, vibration, excitation and ionization cross-sections are used to compute the summed energy-dependent cross-section.  These energy-dependent cross section on each gas are calculated using a finite energy binning, which can be specified by the user, or calculated quickly on-the-fly via an iterative procedure within {\tt PyBoltz}. This procedure is illustrated in the flowchart of Fig~\ref{fig:PyBoltz_Flow}.  The so-determined energy binning is also used to report electron energy distributions after thermalization.  When a physical collision is realized, a gas species from the mixture is selected based on the concentration-weighted, energy-dependent cross sections in the relevant energy bin.  Electron final state kinematics following scattering are drawn using one of a small number of scattering formalisms, which can be specified by the user.  The presently available methods include the anisotropic scattering formalisms of Okhrimovskyy{\em et al.}~\cite{okhrimovskyy2002electron} and Capitelli {\em et al.}~\cite{capitelli2000collision}, as well as simple isotropic scattering.  In the case of Capitelli {\em et al.}~\cite{capitelli2000collision}, the angular distribution is calculated from the provided momentum-transfer and total cross sections at runtime.  For Okhrimovskyy{\em et al.}~\cite{okhrimovskyy2002electron}, the angular distribution parameters are provided, pre-calculated from the cross sections, within the gas data tables.

Before and after each collision the energy, velocity, and position are updated and stored. The drift velocity is calculated per sample, given the total drift distance $Z$ and time $T$, via:
\begin{equation}
v_d = \frac{Z}{T}.
\end{equation}
Diffusion constants are calculated iteratively from the instantaneous electron coordinates per collision [$x_i$,$y_i$,$z_i$,$t_i$] via the equations:
\begin{eqnarray}
D_x = \frac{1}{2}\sum _i^{N_{coll}} \frac{(x_i - x_{i-M})^2} {t_i -t_{i-M}} \times w_i\\
D_y = \frac{1}{2}\sum _i^{N_{coll}} \frac{(y_i - y_{i-M})^2} {t_i -t_{i-M}} \times w_i\\
D_z = \frac{1}{2}\sum _i^{N_{coll}} \frac{(z_i - z_{i-M} - \hat{v}_d (t_i -t_{i-M}))^2}{t_i -t_{i-M}} \times w_i
\end{eqnarray}
From these constants, the conventionally defined $D_L$ and $D_T$ can be extracted, according to:
\begin{eqnarray}
D_L & = & D_z\\
D_T & = & (D_x+D_y)/2
\end{eqnarray}
In the ensemble-averages for $D_{x,y,z}$, the right factor converts the sum into a time-average by weighting according to the time between collisions as $w_i = \frac {t_i - t_{i-1}}{T}$.  The left encodes the mean square distance the electron has migrated over a large number of collisions $M$. This is divided by the time taken for those $M$ collisions to occur. The sum over $i$ then gives a suitable ensemble-average that converges to the diffusion constant.  For accurate convergence the ``decorrelation number'' $M$ must be suitably long that the positions at time $t_i$ are uncorrelated with those at $t_{i-M}$.  The admissible values of $M$ are larger for pure noble gas mixtures than for mixtures with molecular additives, since the latter cool the electrons and suppress their correlations over time.  The decorrelation number as implemented is slightly more complex than the simple picture above, applying several sequential sums at different integer multiples of the  decorrelation distance, once electrons have travelled sufficiently far to reach a steady state behaviour.  The decorrelation parameters can be set by hand, or set to zero in order to be assigned automatically by {\tt PyBoltz}.

It is notable that the expression for diffusion in the $z$ direction, $D_z$, requires prior knowledge of the drift velocity $v_d$ in order be calculated.  This parameter is therefore only determined after the first two samples have been processed in order to estimate $v_d$.  In each sample, the present best estimate of $v_d$, which we label $\hat{v}_d$ is used in the calculation of $D_z$.  Other quantities, for example, the mean electron energy, and the full diffusion tensor including correlations, are also calculated using a method similar to the one described above.  Also accessible within the PyBoltz object are the collision-weighted energy spectrum and information about each individual collision.

\section{ PyBoltz Code Description}
\label{ PyBoltz }
\subsection{Program flow and structure}

As part of the refactorization from {\tt FORTRAN} into more modern languages, the structure of the code has been changed to reflect modular design principles. Program flow is handled by a central {\tt PyBoltz} object, written and compiled in {\tt Cython}.  A user friendly wrapper script {\tt PyBoltzRun} can optionally be used, which makes interactive passing and receiving of parameters more straightforward.  Example scripts are provided for both wrappered and un-wrappered interface modes.  There are independent modules for handling gas data ({\tt Gases}) and Monte Carlo propagators ({\tt Monte}).  All variables have names in natural english and the majority of the code is extensively commented.

The flow of {\tt PyBoltz} follows the flow of {\tt Magboltz}. This starts by setting up the required global constants and values, such as the correlation length, and electron charge and mass, and so on (step 1). {\tt PyBoltz} then estimates the appropriate energy binning for the specified gas mixture (steps 2 and 3). This is done by iteratively choosing an energy binning, calculating cross sections, propagating over a small simulation distances, and testing whether the electron energy spectrum overflows the last bin.  Once the energy scale has been determined, the {\tt Mixer}  object populates all binned cross section arrays appropriately. From the output of the gas mixing object, collision frequencies are extracted and stored in memory as members of the {\tt PyBoltz} object. Finally, an appropriate Monte-Carlo integrator function is called from the {\tt Monte} module to simulate a large number of collisions and to calculate drift properties from simulated collisions (step 4), using the algorithm outlined in the previous section. Outputs are returned as member variables of the {\tt PyBoltz} object, or if using the {\tt PyBoltzRun} wrapper, as named members in {\tt Python} dictionaries.

\subsection{Performance Testing}

\begin{figure}
\begin{centering}
\includegraphics[width=0.5\columnwidth]{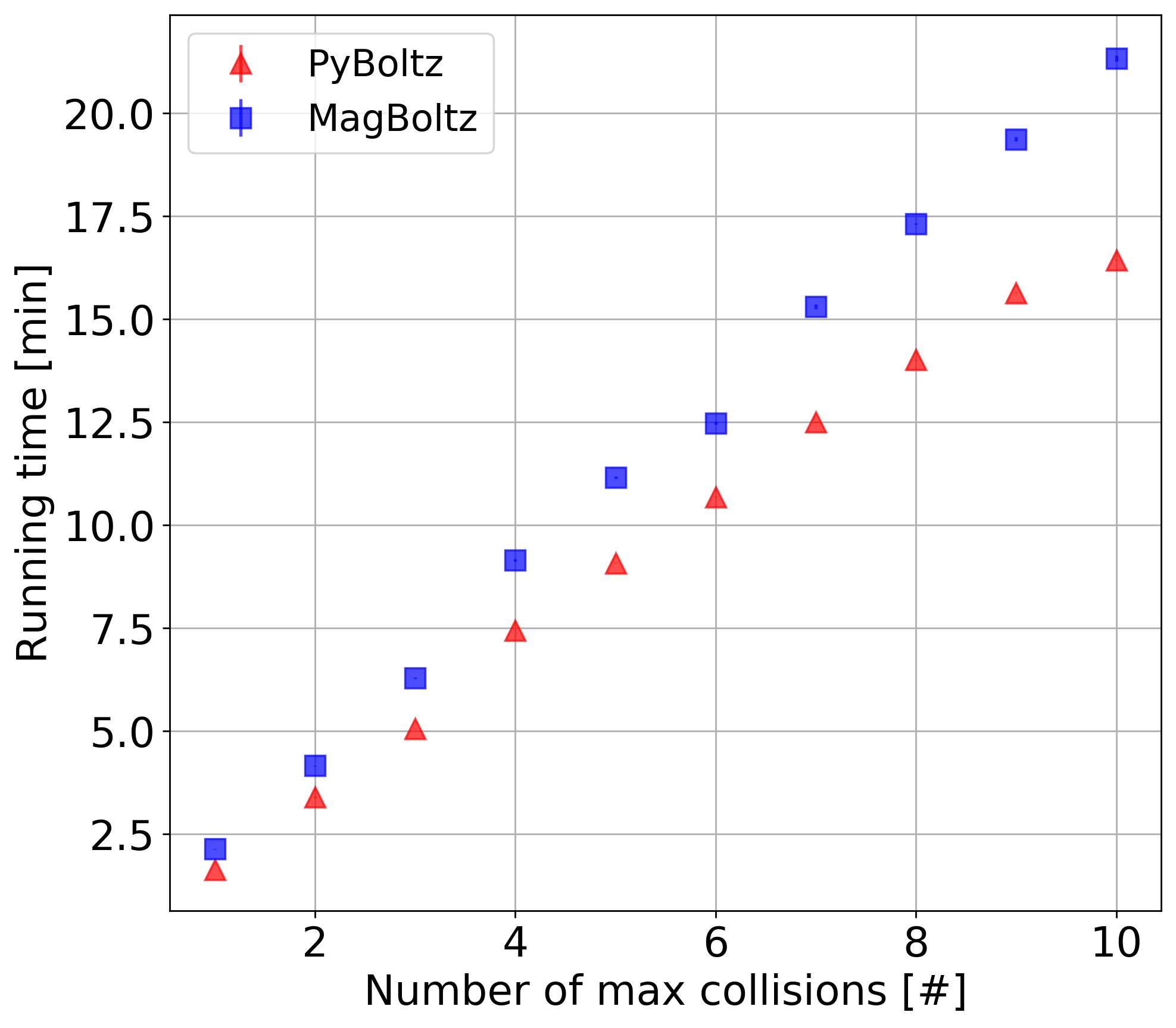}
\par\end{centering}
\caption{ {\tt PyBoltz }  speed comparison to {\tt MagBoltz}. The error on the running times was derived from repeating the measurement 5 times and is on the order of 100ms, which is too small to be visible above.  The collision numbers here are reported in terms of the {\tt MagBoltz} convention, as multiples of $N_{coll}=4\times10^7$.
\label{fig:PyBoltz_Pref}}
\end{figure}

One motivation for choosing {\tt FORTRAN} as a language for scientific computation has been the superior speed of execution afforded by such a low-level language.   There are significant advantages to higher level languages like {\tt Python} (when simple interfaces or interpretive execution are preferred), or C++ (for codes with a complex, modular structure), especially for faster, modern computers, when performance allows.  In the case of {\tt MagBoltz}, the computations for gas mixtures of interest remain intensive, sometimes requiring several hours to scan the parameter points of interest on one CPU. This implies that obtaining optimal code performance is an important requirement when considering potential re-factorizations.

The {\tt Cython} language is a hybrid that effectively compiles {\tt Python}-like code into C, offering much of the flexibility and usability of {\tt Python} with the improved performance of a compiled language. Using {\tt Cython}, {\tt PyBoltz} enjoys the combined benefits of a modular structure, {\tt Python}-like code, and fast execution.  The speeds of the {\tt FORTRAN} and {\tt Python} implementations were directly compared. For this purpose, a test system consisting of CO$_2$ at 1 bar at an electric field of 1000V/cm was chosen.  The results of this performance comparison are shown in Fig.~\ref{fig:PyBoltz_Pref}.  As demonstrated there, the {\tt Cython} implementation outperforms the {\tt FORTRAN} one with speed enhancements at the 20\% level, independent from the number of collisions simulated.  A similar trend has been observed consistently during various cross-checks of the two codes.

Even more important than speed is accuracy, and the two codes have been cross-compared against each other and against data for several systems of interest. We report results of these validations in Sec.~\ref{vali}.

\section{Validation with Transport data}

\label{vali}

In this section we compare the predictions of {\tt PyBoltz} and {\tt MagBoltz} Monte Carlo implementations with data taken in various gas mixtures and experimental configurations. 

In time projection chambers, proportional counters, and other systems employing charge gain using noble gases, molecular additives are commonly used to quench VUV photons that can facilitate electrical breakdowns.  A second property of molecular gases added to noble gases is that they cool the electrons during drift, via efficient transfer of excess energy to various rotational and vibrational degrees of freedom of the additive.  This results in reduced diffusion constants relative to pure noble gases, and tunable drift velocity by up to two orders of magnitude.  A wealth of experimental data exists on these mixtures, and we have picked three model systems with which to compare the accuracy of {\tt PyBoltz} and {\tt MagBoltz}.

A validation data set was chosen from Ref.~\cite{Brockmann:1994gz}, which contains measurements form two well studied gas mixtures: Ar-CH$_4$ (also known as P10), the gas mixture used in the STAR TPC, and Ne-CO$_2$, the base gas mixture used in development of the ALICE experiment. Choosing these mixtures not only gives a robust data comparison but tests the performance of the code with mixed rather than single gases. 
\begin{figure}
\begin{centering}
\includegraphics[width=0.9\columnwidth]{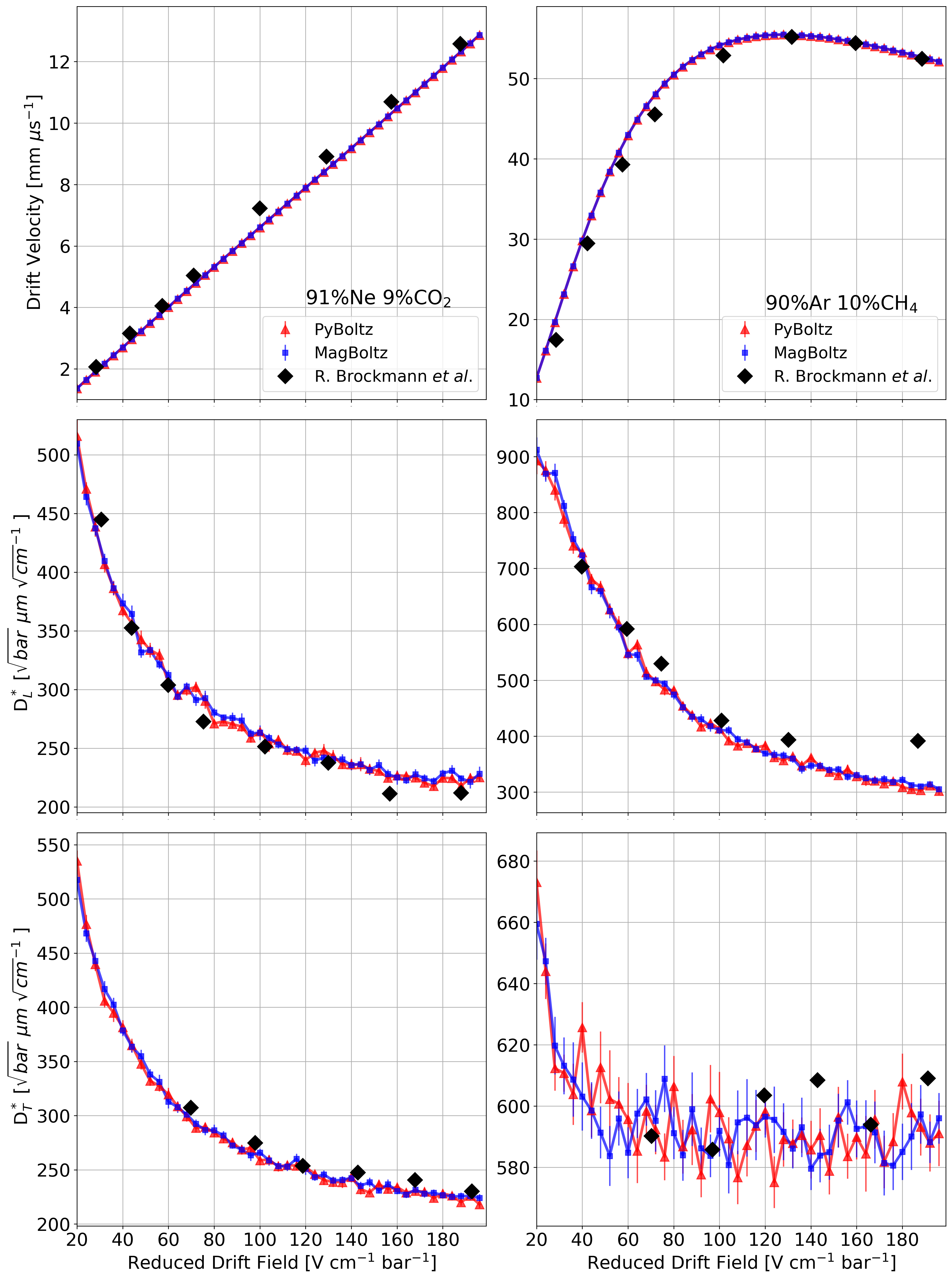}
\par\end{centering}
\caption{Left: The drift velocity, longitudinal and transverse diffusion of 91\%Ne 9\%CO$_2$. Right:The drift velocity, longitudinal and transverse diffusion of 90\%Ar 10\%CH$_4$. All calculations are at standard temperature and pressure. Data sets were taken from Ref~\cite{Brockmann:1994gz}.
\label{fig:PyBoltzDataComp}}
\end{figure}

Simulations with both {\tt MagBoltz} and  {\tt PyBoltz} were executed at 1~bar pressure over the range of electric fields (20-200 V/cm). This range was chosen because most gaseous TPCs operate in this region of reduced field (V/cm/bar).  Irrespective from the operating pressure, scaling of drift-diffusion parameters with reduced field is generally satisfied in this regime~\cite{Gonzalez-Diaz:2017gxo}, and so the results can be scaled to other pressures and drift fields using standard methods.  In order to achieve accurate results the number of collisions was set to $4\times10^{8}$ divided across ten samples, which was found sufficient to achieve convergence without an excessive run time. The gas mixtures were set to 90\% argon with 10\% CH$_4$ and 91\% Ne with 9\% CO$_2$, to match the reported fractions in the Ref.~\cite{Brockmann:1994gz}. Both simulations agree to high degree of accuracy and both match the data sets within simulation error bars. The strong agreement in $D_L$, $D_T$ and $V_z$ can be seen in Fig.~\ref{fig:PyBoltzDataComp}.

\begin{figure}
\begin{centering}
\includegraphics[width=0.5\columnwidth]{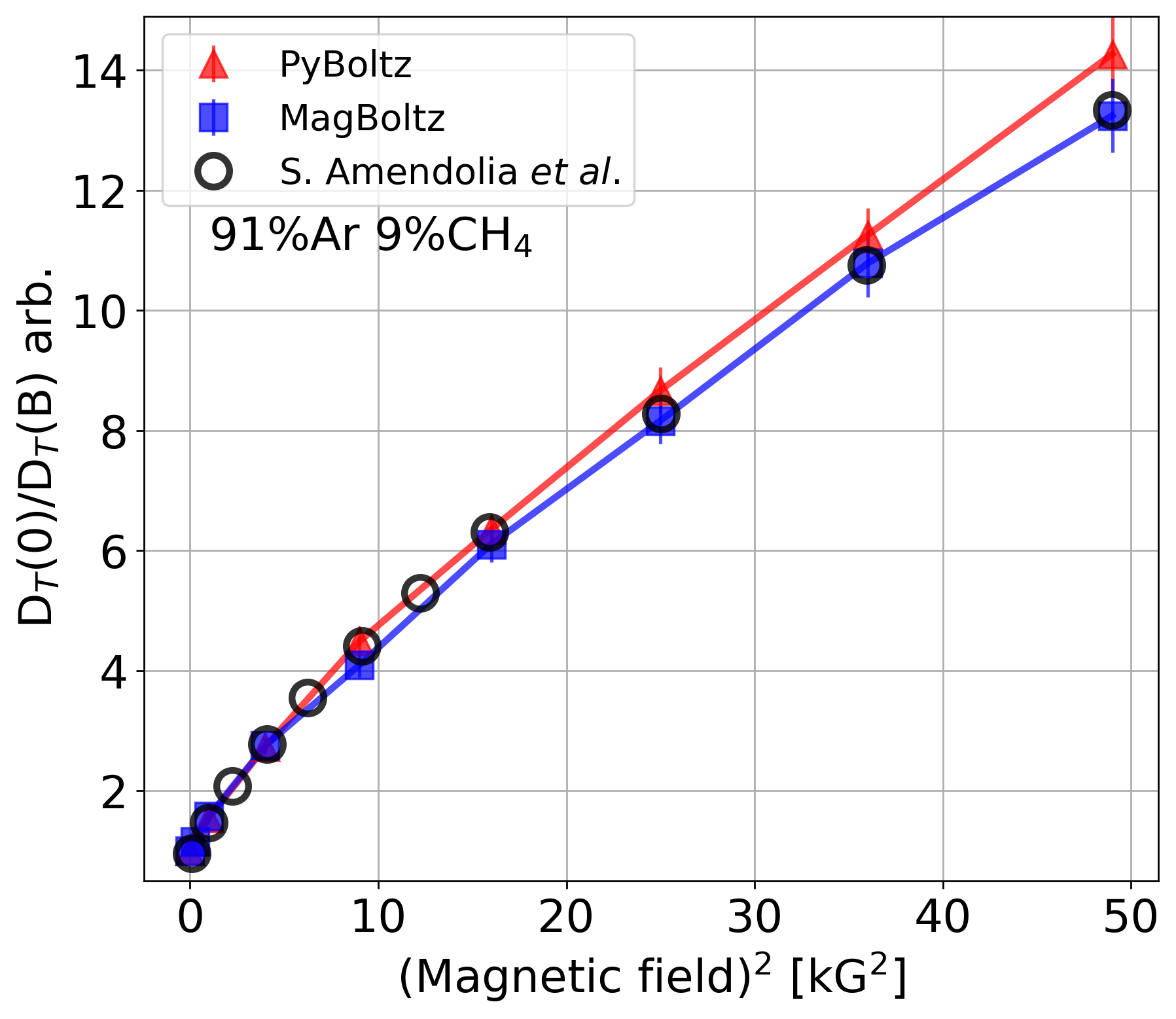}
\par\end{centering}
\caption{Validation of B-field suppression of transverse diffusion in a $91\%$Ar-$9\%$CH$_4$ gas mixture, {\tt MagBoltz} and {\tt PyBoltz} compared to data of Ref~\cite{amendolia1986dependence} .}
\label{fig:PyBoltzBfield}
\end{figure}

For the case of electron transport in magnetic fields, data from Ref~\cite{amendolia1986dependence} were used as a benchmark. When a magnetic field is applied parallel or anti-parallel to the electric field, the diffusion in the transverse direction is reduced due to cyclotron motion of drifting electrons which prevents their lateral spreading~\cite{Nygren:1978rx}. Reduction of the transverse diffusion is critical for retaining signal information when drifting electrons over a long distance in a TPC, and we consider this case here for illustration. The ratio $\frac{D_T(0)}{D_T(B)}$  is used as a measure to quantify B-field assisted diffusion suppression. Simulations were run in a similar manner to the no B-field simulations, but using an electric field of 115 V/cm and a magnetic field in the $z$ direction between 0 and 0.7~Tesla. The gas was set to 91\%argon with 9\%CH$_4$ to match the measurement conditions of Ref~\cite{amendolia1986dependence}. 
The transverse diffusion in $\frac{cm^2}{s}$ was taken from both programs and the ratio $\frac{D_T(0)}{D_T(B)}$ calculated for each field point. Once again, the {\tt PyBoltz} and {\tt MagBoltz} calculations are consistent with each other and with measured data within statistical uncertainties, as shown in Fig~\ref{fig:PyBoltzBfield} left.

\begin{figure}
\begin{centering}
\includegraphics[width=1\columnwidth]{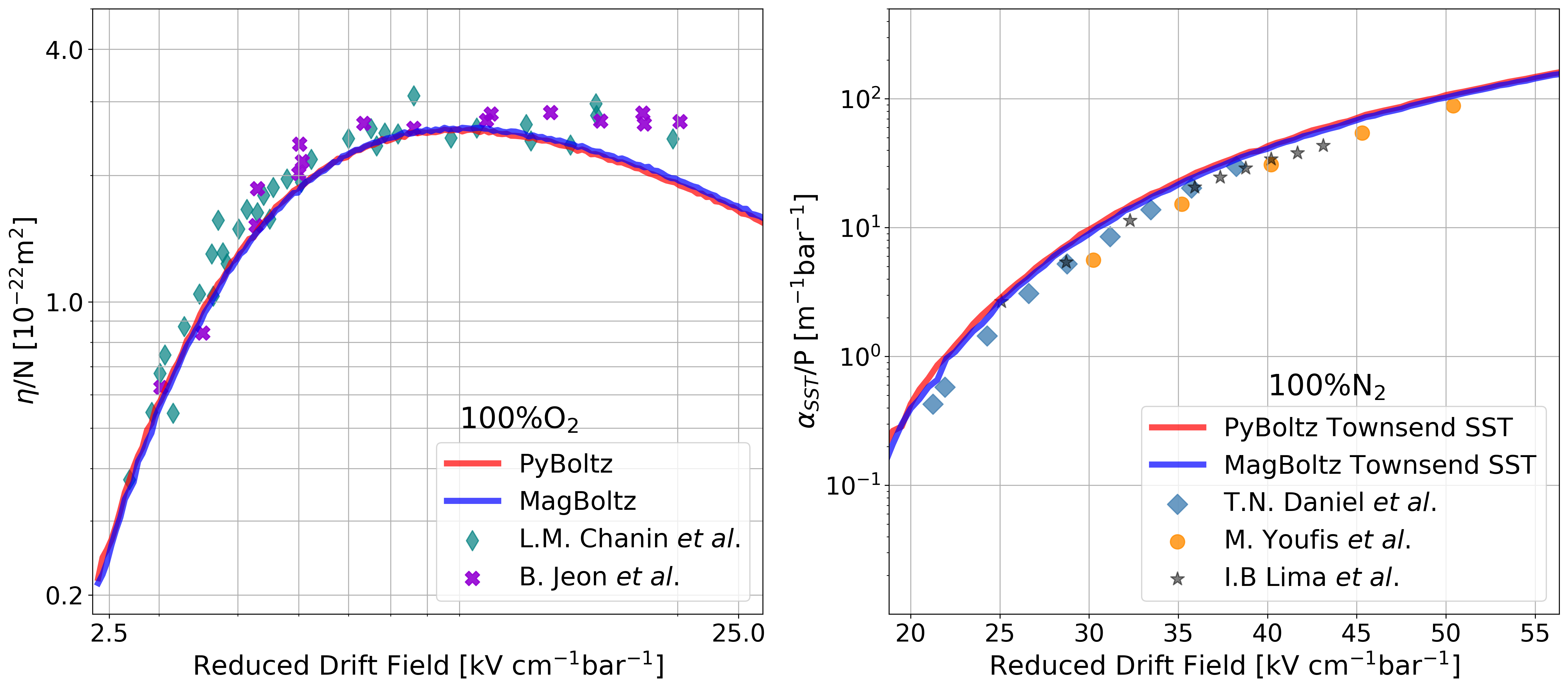}
\par\end{centering}
\caption{Left: The reduced attachment rate in pure oxygen generated by {\tt MagBoltz} and {\tt PyBoltz} and compared to data from \cite{chanin1962measurements,jeon1998measurement}. Right: Comparison of the Townsend coefficient generated by {\tt MagBoltz} and {\tt PyBoltz} with pure nitrogen data from \cite{lima2012experimental, daniel1970spatial, yousfi2009electron} obtained at standard temperature and pressure. }
\label{fig:town-att}
\end{figure}

The electron attachment coefficient $\eta$ can be calculated, for mixtures containing attaching gases, based on tables of electron attachment cross sections.  A particular interesting example is that of oxygen which exhibits two and three-body attachment, both accounted for within {\tt PyBoltz} and {\tt MagBoltz}.  Simulations are compared to data from Refs.~\cite{chanin1962measurements,jeon1998measurement} are shown in Fig~\ref{fig:town-att} left.

{\tt PyBoltz} and {\tt MagBoltz} are both equipped for the calculation of the Townsend multiplication coefficient $\alpha$. This calculation is initiated when the absolute difference of the ionization and attachment rate is greater than a threshold, by default set to 30 electrons per centimeter, in reduced units.  This supplementary calculation utilizes a separate Monte Carlo function that tracks both the primary and secondary electrons.  There are two approaches to the calculation of Townsend coefficients, appropriate for steady state ($\alpha_{SST}$) or pulsed-discharges ($\alpha_{PT}$), respectively. The distinction and relationship between these coefficients is discussed in detail in~\cite{crompton1967comments}, and they become equivalent at low reduced fields, in practice below around 25 kV/cm/bar at room temperature, in many gases. Calculation of $\alpha_{PT}$ proceeds via tracking an electron through evenly spaced time planes, extracting the average production of electrons per unit time (and dividing by the drift velocity); whereas calculation of $\alpha_{SST}$ follows a similar approach but with evenly spaced distance steps (without dividing by the drift velocity).  {\tt PyBoltz} can accommodate both mixtures where Penning transfer reactions between components are important, as well as pure gases where this effect is sub-dominant~\cite{csahin2010penning,ruiz2015modelling,lima2012experimental}. In Fig.~\ref{fig:town-att}, right, we compare the measured $\alpha_{SST}$ values in pure nitrogen gas~\cite{lima2012experimental, daniel1970spatial, yousfi2009electron}) to {\tt PyBoltz} and {\tt MagBoltz} calculations, finding good consistency between both codes and data.

\section{Conclusion}
\label{Conclusion}
The widely used and influential gaseous detector simulation code {\tt Magboltz} has been refactorized into {\tt PyBoltz}, a {\tt Cython} based code that has a modular structure allowing for improved flexibility and potentially a widened developer base.  {\tt PyBoltz} has demonstrated performance that is comparable to {\tt MagBoltz} in both computation time and precision, with reproduction of results calculated with { \tt MagBoltz} to a high degree of accuracy.  The extendability of a modular, {\tt Python}-based code has also opened up the potential to develop extension packages for new applications.  We provide two examples in Appendix A of this work.  We hope that in the form of {\tt PyBoltz}, the invaluable and seminal contributions made by {\tt MagBoltz} to the field of gaseous detectors will continue as the code-base is embraced and extended by further generations of detector physicists.

\section*{Appendix A: Examples of Extended PyBoltz Applications}

One goal of developing {\tt PyBoltz} has been to modularize the code and enable extendability.  We present two examples in this section: first, the implementation of tunable cross sections with Modified Effective Range Theory (MERT); and second a wide exploration of properties of high pressure noble gases with additives (the ``Plus Anything 2.0'' program).

\subsection{Tunable Cross Sections with Modified Effective Range Theory}
Modified Effective Range Theory (MERT) utilizes the phase-shift representation of scattering cross sections in order to parameterize scattering behaviours that are consistent with angular momentum conservation in quantum mechanics. The original MERT method was developed by O'Malley \cite{o1963extrapolation}, is explained in detail by Raju in Ref.~\cite{raju2005gaseous}, and has been used to obtain phenomenological fits to various low energy cross sections in various more recent works.  While the cross sections can be fitted with electron beam data as in Ref.~\cite{kurokawa2011high}, it is also common to attempt to fit the cross section, using electron drift parameters as demonstrated in Refs.~\cite{o1963extrapolation, HunterLow1988, pack1992longitudinal}.

\begin{figure}[t]
\begin{centering}
\includegraphics[width=0.85\columnwidth]{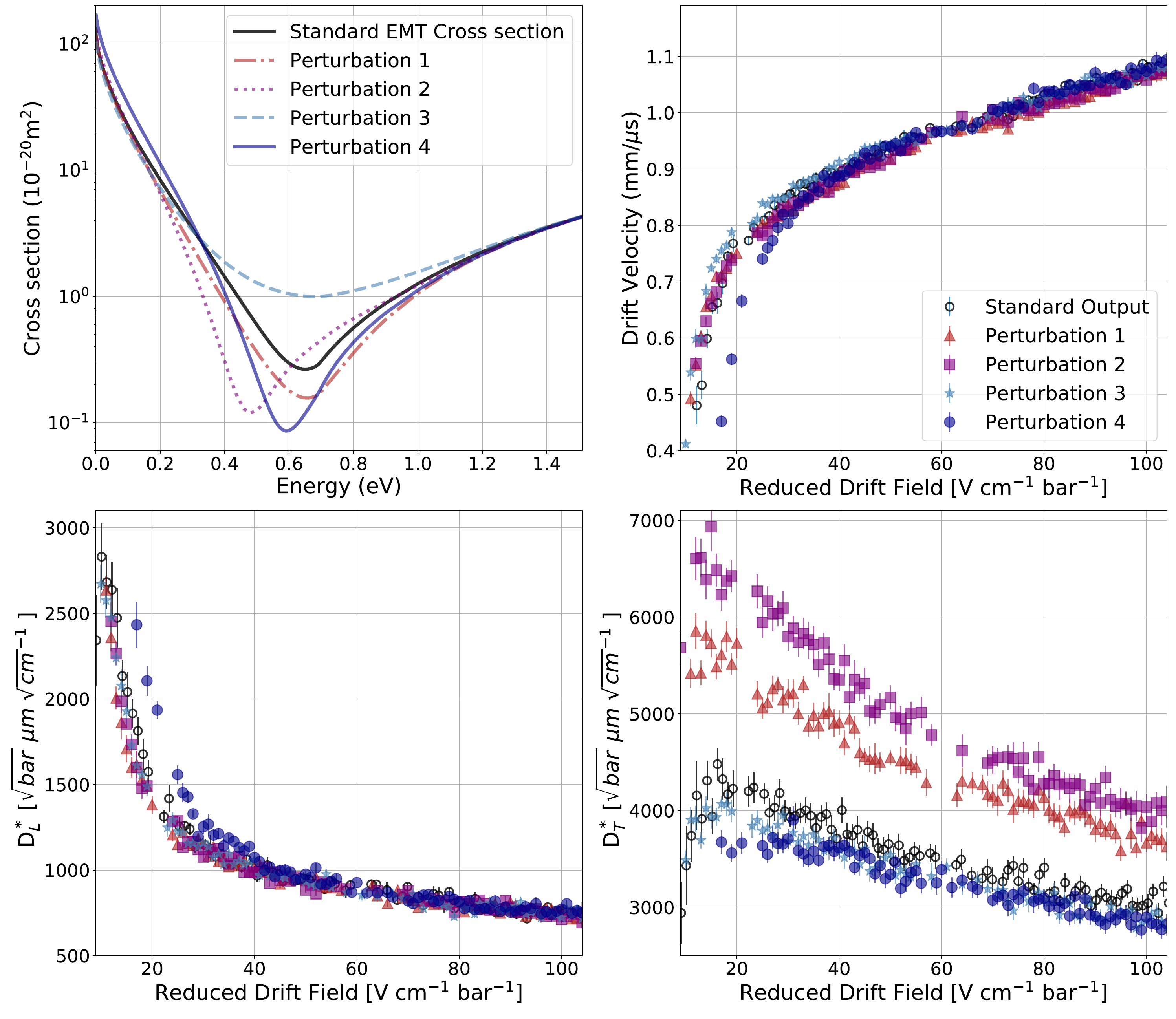}
\par\end{centering}
\caption{Top left: Default momentum transfer cross section in {\tt PyBoltz} alongside different parametrizations consistent with the MERT formalism. Top right, bottom left, and bottom right display the effect of changing cross sections on drift velocity, longitudinal, and transverse diffusion respectively. 
\label{fig:PyBoltzMERT}}
\end{figure}

{\tt PyBoltz} allows for scattering parametrizations such as MERT to be directly implemented into its modular structure. Our recent work has focused on testing perturbations of the xenon cross in the vicinity of its pronounced Ramsauer minimum, and will be the subject of future publications. The second order MERT formalism in \cite{HunterLow1988} is presently available within  {\tt PyBoltz}, with modern extensions such as MERT5/6 from \cite{kurokawa2011high} being implemented for future releases.
The zeroth and $l$'th MERT phase shifts are described by:
\begin{eqnarray}
\tan(\eta_0) &&= -Ak[1+\frac{4\alpha}{3a_0}k^2ln(3a_0)]-\frac{\pi \alpha}{3a_0}k^2 + Dk^3+Fk^4\\
\tan(\eta_1) &&= \frac{\pi}{15 a_0} \alpha k^2 [1 - {\Big (\frac{\varepsilon}{\varepsilon_1} \Big )}^{\frac{1}{2}}]\\
\tan(\eta_l) &&= \pi \alpha k^2 / [(2l+3)(2l+1)(2l-1)a_0]
\end{eqnarray}
where $A$, $D$, $F$, $\varepsilon_1$ are the adjustable MERT parameters, $\alpha$ is the polarizability of the atom (27.292$a_0^3$ for xenon), $a_0$ is the Bohr radius, $k$ is the electron wavenumber that is related to the energy via $\epsilon = 13.605(ka_0)^2$, $l$ is the angular momentum quantum number, and $\eta_0$, $\eta_1$ represent the zeroth and first phase shift respectively.
With this parameterization the momentum-transfer and total cross sections are given by:
\begin{eqnarray}
\sigma_m  &&= \frac{4\pi a_0^2}{k^2} \sum_{l=0}^{\infty} (l+1)\sin^2(\eta_l-\eta_{l+1}),\\
\sigma_t  &&= \frac{4\pi a_0^2}{k^2} \sum_{l=0}^{\infty} (2l+1)\sin^2(\eta_l).
\end{eqnarray}

Given a set of MERT parameters, the elastic and momentum-transfer cross sections are calculated in {\tt PyBoltz} in a consistent manner prior to swarm evolution. The cross section for xenon are only modified below 1 eV where MERT is applicable, and merged smoothly into the standard Biagi cross section at high energy, as described in Refs~\cite{HunterLow1988,kurokawa2011high}. 

Simulating data sets in reduced field space with various cross sections allows to fit the cross sections with experimental data. A grid search over various cross sections can then be used to re-fit cross sections to data. An example showing the drift properties obtained with various MERT parameters is presented in Fig.~\ref{fig:PyBoltzMERT}.

\begin{figure}[t!]
\begin{centering}
\includegraphics[width=0.99\columnwidth]{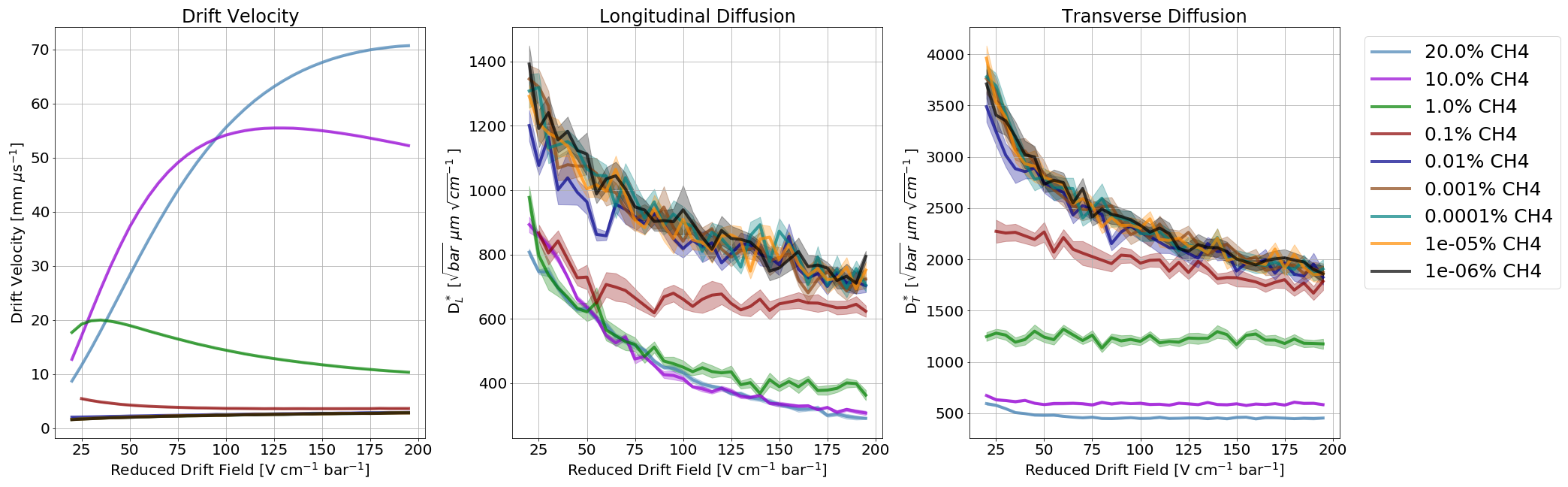}
\includegraphics[width=0.99\columnwidth]{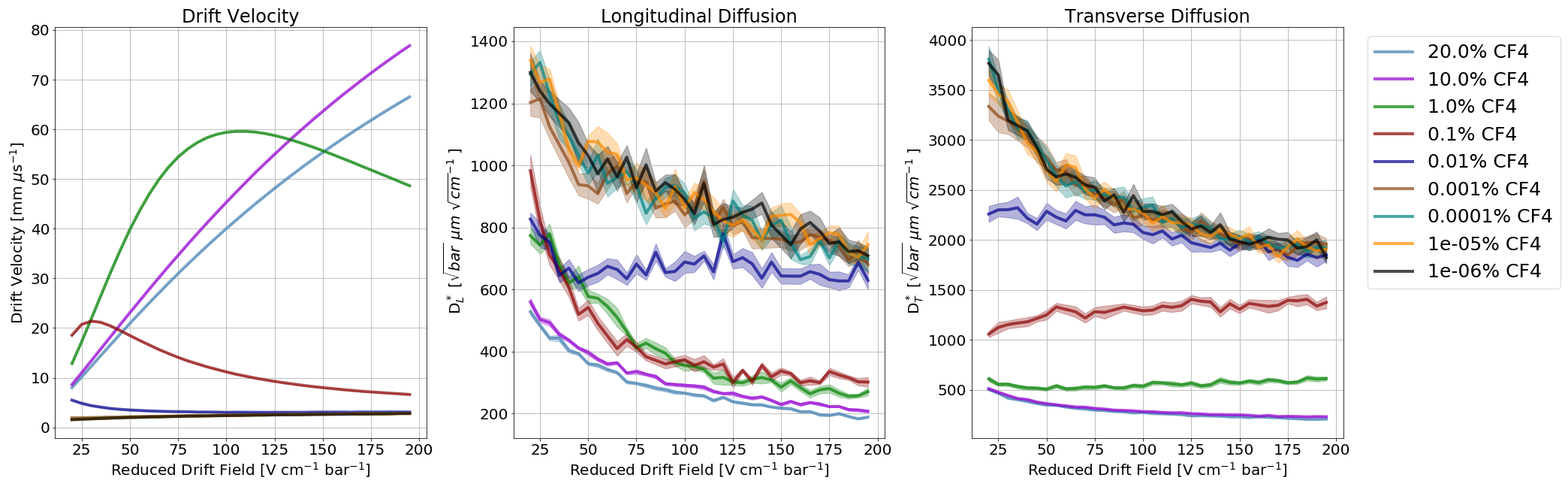}
\includegraphics[width=0.99\columnwidth]{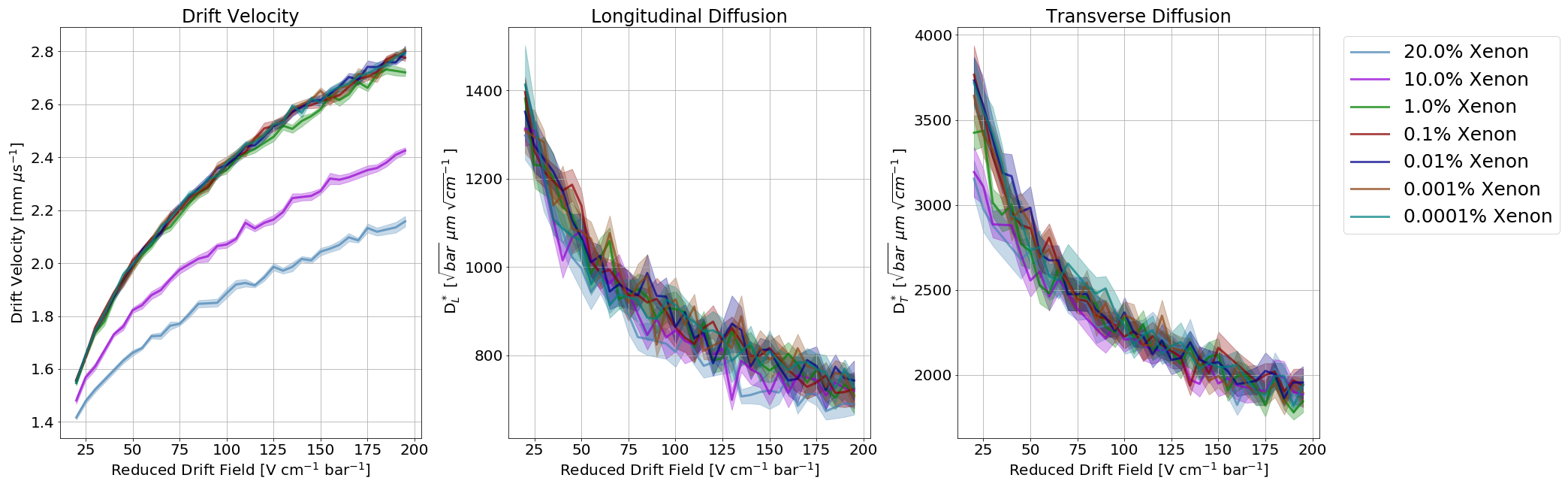}
\par\end{centering}
\caption{Example plots from the ``Argon Plus Anything'' project using PyBoltz. This example uses argon with various concentrations of CH$_4$, CF$_4$ and xenon. We show here drift velocity, longitudinal and transverse diffusion coefficients. \label{fig:test2}}
\end{figure}

\subsection{Argon or Xenon ``Plus Anything'' 2.0}
The {\tt PyBoltz} code can be easily parallelized to run using distributed computing resources.  In the past we explored combinations of xenon ``Plus Anything'' gas mixtures, studying admixtures between 10$^{-6}$\% to 20\% in xenon at 10 bar and room temperature to survey the diffusion reducing properties of molecular and light noble gases~\cite{XePa}.  Using our new package we have re-calculated these data points, which can be found at Ref.~\cite{GitHubPlusAny}.

A recent application of high pressure gas mixtures concerns the DUNE MultiPurpose Detector~\cite{Martin-Albo:2016tfh}.  R\&D is required to optimize the gas mixture for this device, which is under way at various institutions, with the goal of enabling fast, low-diffusion, quenched and HV-stable drift of electrons in a gas of predominantly 10 bar argon.  We have surveyed ``Argon Plus Anything'' gas mixtures at concentrations between 10$^{-6}$\% to 20\% of CF$_4$, CH$_4$, DME, $^4$He, H$_2$, isobutane, krypton, neon, propane and xenon, among others. Studying mean electron energy, drift velocity, longitudinal and transverse diffusion and attachment.  The simulations were performed with electric fields between 20-200V/cm at a pressure of 1 bar, with simple extrapolation to any other field by the well known universality of transport parameters at a given reduced field $E/N$ \cite{sauli2014gaseous} (this universality was verified in preliminary studies with a subset of mixtures) where $E$ is the electric field and $N$ is the molecular or atomic number density.  A compilation of the simulation results are available at~\cite{GitHubPlusAny}, with a few examples shown in Fig.~\ref{fig:test2}.

\clearpage

\section*{Appendix B: Default Parameters in PyBoltz}
\begin{center}

\begin{turn}{90}
\begin{tabular}{ |l | c |l| }
  \hline			
  {\bf Parameter} & {\bf Default}  & {\bf Description} \\
  \hline
  {\tt Gases} & {\tt ["NEON", "CO2"]} & List of named gases to use. \\
  {\tt Fractions} & \tt{ [90,10] }& Percentages of each gas to use\\
  {\tt Max\_Collisions} & {\tt 4$\times 10^7$ }& Total number of collisions to simulate \\
  {\tt Num\_Samples}  & {\tt 10 }& Number of points to statistically sample drift parameters \\
  {\tt Temperature\_C} &{\tt  23 }& Gas temperature in Celcius \\
  {\tt Pressure\_Torr} & {\tt 750.062 }& Gas pressure in Torr \\
  {\tt BField\_Tesla} & {\tt 0} & Applied magnetic field in Tesla \\
  {\tt BField\_Angle} & {\tt 0} & Angle of magnetic to electric field \\
  {\tt Enable\_Penning} & {\tt 0} & Switch on or off Penning transfer effects \\
  {\tt Enable\_Thermal\_Motion} & {\tt 1 }& Switch on or off thermal motion of the gas \\
  {\tt Console\_Output\_Flag} &{\tt  0 }& Print output messages to console \\
  {\tt Decor\_Colls} & {\tt 0 }& Distance swarm must evolve to begin counting $^{\dagger }$ \\
  {\tt Decor\_Lookbacks} & {\tt 0} & How many points to correlate electron with itself $^{\dagger}$\\
  {\tt Decor\_Step} &{\tt  0} & Decorrelation number parameter M over which correlations vanish  $^{\dagger }$\\
  {\tt Max\_Electron\_Energy} & {\tt 0} & Upper limit of binned energy spectrum  $^{\dagger }$\\
 {\tt Steady\_State\_Threshold} & {\tt 40} & Steady State Townsend threshold (see text)\\
 {\tt Which\_Angular\_Model} & {\tt 2} & Choose one of the three angular distribution models \\
  {\tt Random\_Seed} & {\tt 54217137} & Seed value for the random number generator\\
  \hline			
  & & $^{\dagger}$ implies automatically estimated if zero \\
  \hline			

\end{tabular}
\end{turn}
\end{center}

\section*{Acknowledgement}
We would like to thank Roxanne Guenette and the Harvard FAS Research Computing center for the use of the Odyssey cluster, and members of the NEXT collaboration including Neus Lopez March and Ryan Felkai for illuminating conversations.  The UTA group acknowledges support from the Department of Energy under Early Career Award number DE-SC0019054, and by Department of Energy Award DE-SC0019223. DGD acknowledges the Ramon y Cajal program (Spain) under contract number RYC-2015- 18820.

\bibliography{main}
\bibliographystyle{elsarticle-num}

\end{document}